\documentclass{sigchi-ext}
\usepackage[T1]{fontenc}
\usepackage{textcomp}
\usepackage[scaled=.92]{helvet} 
\usepackage{graphicx} 
\usepackage{balance}  
\usepackage{booktabs} 
\usepackage{ccicons}  
\usepackage{ragged2e} 
\usepackage{subfigure}
\copyrightinfo{Copyright is held by the author/owner. Permission to make digital or hard copies of all or part of this work for personal or classroom use is granted without fee. Poster presented at the 14th Symposium on Usable Privacy and Security (SOUPS 2018).}

\def\plaintitle{SIGCHI Extended Abstracts Sample File: Note Initial
  Caps} 
\def\emptyauthor{}
\def\plainkeywords{Authors' choice; of terms; separated; by
  semicolons; include commas, within terms only; required.}

\title{Visual Analysis of Photo Policy Misconfigurations Using Treemaps}

\numberofauthors{2}
\author{%
  \alignauthor{%
    \textbf{Yousra Javed}\\
    \affaddr{Illinois State University} \\
    \affaddr{School of Information Technology} \\
    \affaddr{Normal, IL 61790 USA}\\
    \email{yjaved@ilstu.edu} } 
   \\\\\\ \alignauthor{%
    \textbf{Mohamed Shehab}\\
    \affaddr{University of North Carolina Charlotte} \\
    \affaddr{Department of Software \& Information Systems} \\
    \affaddr{Charlotte, NC 28223 USA}\\
    \email{mshehab@uncc.edu} }} 

\definecolor{linkColor}{RGB}{6,125,233}
\hypersetup{%
  pdftitle={\plaintitle},
  pdfauthor={\emptyauthor},
  pdfkeywords={\plainkeywords},
  bookmarksnumbered,
  pdfstartview={FitH},
  colorlinks,
  citecolor=black,
  filecolor=black,
  linkcolor=black,
  urlcolor=linkColor,
  breaklinks=true,
}

\begin{document}

\maketitle

\RaggedRight{} 


\begin{abstract}
Online photo privacy is a major concern for social media users. Numerous visualization tools have been proposed to help the users easily compose and understand policies on social networks. However, these tools do not incorporate the ability to quickly identify and fix unintended photo sharing. We propose a tool that displays the photo albums w.r.t their policy misconfigurations using a Treemap visualization. 
\end{abstract}
\section{Introduction}
Photos are one of the most frequently shared content on social media \cite{mmtStudy}. Braunstein et al. \cite{braunstein2011indirect} conducted a study using indirect content privacy surveys. Their results indicated the users' privacy concern for their online photos. Being non-experts in access control---the users frequently end up giving access to unintended audience. 
Unintended sharing corresponds to both over-sharing or under-sharing of information. This suggests the need for techniques that can enable a user to quickly figure out the unintended audience accessing a photo and fix it.

Numerous visualization tools have been proposed to help the users easily compose and understand policies on social networks. PViz is a tool proposed by Mazzia et al. \cite{Pviz} which lets the user understand the visibility of her profile according to various friend groups. Similarly, Anwar et al. \cite{Anwar} proposed a visualization tool for privacy settings. The social network is displayed as a graph, and mousing over a node in the graph indicates what the person can access in the current user's profile. However, these tools do not incorporate the ability to quickly identify and fix unintended photo sharing. 

A Treemap \cite{Treemaps} displays hierarchical (tree-structured) data as a set of nested rectangles. Each branch of the tree is given a rectangle, which is then tiled with smaller rectangles representing sub-branches. A leaf node's rectangle has an area proportional to a specified dimension of the data. When the color and size dimensions are correlated in some way with the tree structure, one can often easily see patterns that would be difficult to spot in other ways, such as if a certain color is particularly relevant. A second advantage of Treemaps is that, by construction, they make efficient use of space. As a result, they can legibly display thousands of items on the screen simultaneously.

We propose a tool that displays the photo albums w.r.t their policy misconfigurations using a Treemap visualization. We developed a prototype for Facebook by extracting the privacy settings of photo albums through the Facebook API. 

\section{Potential Policy Misconfigurations}
\begin{figure*}[!ht]
 \subfigure[\small{Level 1- Album Policy Groups}]{
  \includegraphics[scale=0.45]{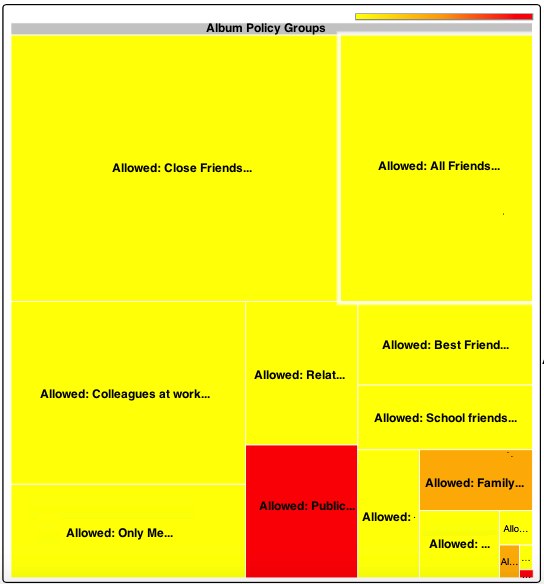}
   \label{fig:level1}
   }
 \subfigure[\small{Level 2 - Albums Inside an Album Policy Group}]{
\includegraphics [scale =0.45]{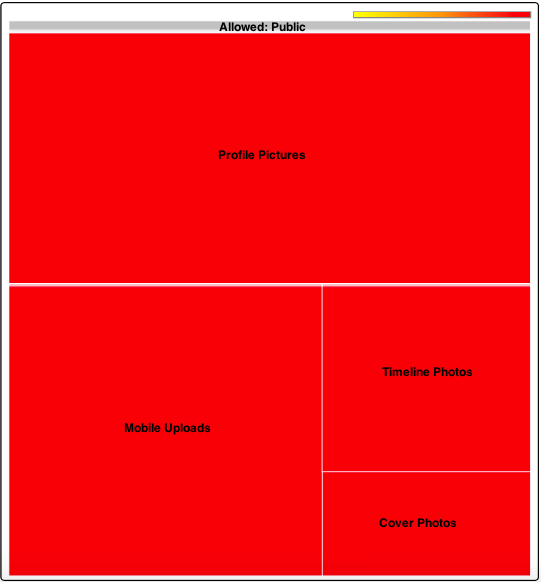}
 \label{fig:level2}
   }
 \caption{Treemap visualization of photo albums w.r.t their policy misconfigurations} \label{fig:Treemap}
\end{figure*}

We use the following set of misconfigurations proposed by Javed et al. \cite{javed2013access}. Each misconfiguration is assigned a sensitivity weight on a scale from 0 to 1 based on the extent of privacy leak that it can cause (Table \ref{tab:MSen}). 

\begin{table}[htbp]
\caption{Misconfiguration sensitivities} \label{tab:MSen}
\small
\begin{center}
\begin{tabular}{|l|c|}
\hline
\textbf{\small{Misconfiguration}}&\textbf{\small{Sensitivity}}\\
\hline
M1. A friend list having no friends is allowed or  & 0.1 \\
 denied &\\
\hline
M2. One or more friends exist in both the & 0.2 \\
allowed and denied fields&\\
\hline
M3. There are common friends between & 0.4 \\
friend lists &\\
\hline
M4. A friend has been denied explicitly in this & 0.6\\
album, but is allowed in other albums & \\
\hline
M5. An automatically created friend list which  & 0.8\\
updates without user concern has been used &  \\
\hline
M6. The album is visible to people outside the  & 1 \\
friend network &\\
\hline
\end{tabular}
\end{center}
\end{table}
\begin{enumerate}
\item[M1] An empty friend list has been allowed or denied. For example, if the user allows access to a smart list which contained friends initially, but became empty over the course of time when the respective friends changed their work, education or hometown.
\item[M2] A friend has been explicitly denied access to this photo album, but is allowed access to other albums. This misconfiguration can arise due to the existence of a friend who has been given access to most albums but is denied access to a specific album by adding their name in the denied set of users. The user can have a false perception that this friend has been denied access to all the other albums as well. 
\item[M3] There are common friends between friend lists. For example, Facebook gives access to the union of all the friends/friend lists included in ``Allowed'' field. However, during policy composition, the user might want to allow access to only those friends who exist in all the friend lists in the ``Allowed'' field, leading to over-sharing of information.
\item[M4] One or more friends exist in both the allowed and denied fields. In Facebook access control, the \textit{deny} permission takes precedence over the \textit{allow} permission. Therefore, this misconfiguration intends to notify the user of unintended denial of access to a friend.
\item[M5] A smart list, which updates automatically without user consent, has been used. If a user grants access to a smart list, there is a potential chance of sharing information with unintended audience because the user does not control the list. 
\item[M6] The photo album is visible to people outside the friend network. This scenario is possible if the chosen privacy settings are Friends of Friends, Friends and Networks, or Public.
\end{enumerate}
\section{Treemap Visualization}
\begin{figure}[ht]
\begin{center}
\includegraphics[scale=.41]{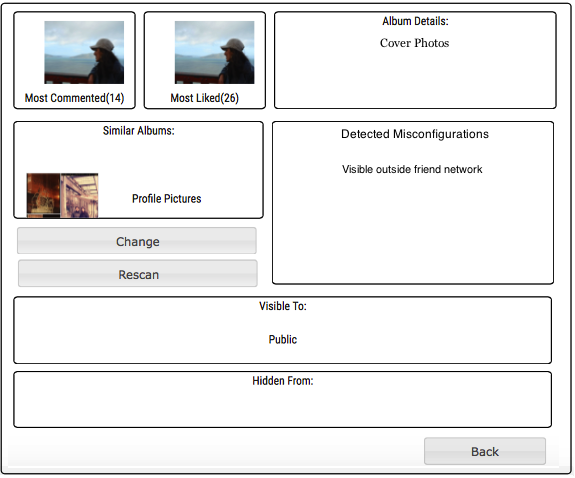}

\end{center}
\caption{Album Summary Screen}\label{fig:albumDetails}
\end{figure}

We use a Treemap to visualize the photo albums w.r.t their policy misconfigurations. 
\subsection{Treemap Levels}
There are two levels in our Treemap hierarchy.
\begin{itemize}
\item Level 1 - The first tree level displays the albums grouped by their policies. Each rectangle in this level corresponds to a distinct policy. The size of the rectangle is proportional to the number of albums having this policy. The color of the rectangle shows the sensitivity of the misconfiguration detected in the underlying policy, where red represents the highest sensitivity and yellow represents the lowest sensitivity (Figure \ref{fig:level1}).
\item Level 2 - The second level displays all the albums inside an album policy group. Each rectangle corresponds to a single album. The size of the rectangle is proportional to the number of photos inside the album. The color of each rectangle is the same as that of the parent rectangle in level 1 and shows the sensitivity of the misconfiguration detected in the album. Figure \ref{fig:level2} shows the albums inside the policy group ``public". Clicking on a rectangle in this level takes the user to the album's details screen. This screen displays a summary of the album consisting of its description, policy, detected misconfigurations, most liked and commented photos, and similar albums. There is also an option to fix the album policy and rescan it (Figure \ref{fig:albumDetails}).
\end{itemize}

\section{Pilot Study}
In order to evaluate our prototype, we are designing a pilot study that answers the following questions:
\begin{enumerate}
\item Compared to the existing privacy configurations interfaces on social media, is the Treemap visualization more effective w.r.t users' speed and accuracy of identifying and fixing album policy misconfigurations? 
\item What is the users' subjective satisfaction of our prototype based on the System Usability Scale (SUS) \cite{sus}?
\item Can we use eye-gaze tracking to analyze user attention towards our prototype? 
\end{enumerate}
\bibliographystyle{SIGCHI-Reference-Format}
\bibliography{sigproc}  

\end{document}